\documentclass[11pt]{article}
\usepackage{times}
\usepackage{geometry}
\usepackage[utf8]{inputenc}
\usepackage{enumitem,amssymb}
\usepackage{ragged2e}
\justifying
\newlist{thematic}{itemize}{8}
\setlist[thematic]{label=$\square$}
\usepackage{pifont}
\usepackage{fancyhdr}
\usepackage[USenglish]{babel}
\usepackage[utf8]{inputenc}
\usepackage[T1]{fontenc}
\usepackage{ulem}
\usepackage{epsfig}
\usepackage{wrapfig}
\usepackage{color}

\usepackage[vflt]{floatflt}
\usepackage{longtable}
\usepackage{caption}
\usepackage{jheppub}   
\usepackage[dvipsnames,svgnames,x11names]{xcolor}
\usepackage[all]{hypcap} 
\usepackage{amssymb,amsmath}
\usepackage{longtable}
\usepackage{amssymb}
\usepackage{amsmath, xspace}
\usepackage{graphics,graphicx} 
\usepackage{rotating}
\usepackage{color}
\usepackage{xcolor}
\usepackage{multirow}
\usepackage{subfigure}
\usepackage{sectsty}
\usepackage{comment}
\usepackage[outercaption]{sidecap}
\sidecaptionvpos{figure}{c}

\newcommand{\cc}[1]{{\textcolor{blue}{#1}}} 

\definecolor{DarkGreen}{rgb}{0.0, 0.3, 0.0}
\definecolor{purple}{rgb}{0.5, 0.0, 0.5}
\definecolor{red}{rgb}{1, 0.0, 0.0}
\definecolor{green}{rgb}{0, 1.0, 0.0}

\newcommand{\Moon}{$M_{Moon}$}                          

\newcommand{\apjl}{\rm ApJL}
\newcommand{\apjs}{\rm ApJS}

\newcommand{\aap}{\rm A$\&$A}

\def\ao{{\it Appl.~Opt.}}           

\def\3he{$^3{\rm He}$}
\def\arcdeg{\hbox{$^\circ$}}

\def\lsim{\mathrel{\lower2.5pt\vbox{\lineskip=0pt\baselineskip=0pt
           \hbox{$<$}\hbox{$\sim$}}}}

\def\gsim{\mathrel{\lower2.5pt\vbox{\lineskip=0pt\baselineskip=0pt
           \hbox{$>$}\hbox{$\sim$}}}}

\begin{document}
\raggedright
\huge
The role of supernova remnants for the emergence of pre-biotic chemistry in molecular clouds.  
\linebreak
\bigskip
\normalsize

\textbf{Authors:} 
Giuliana Cosentino (cosentino@iram.fr; Institute de Radioastronomie Millimétrique, IRAM, France),
Izaskun Jim\'enez-Serra (Centro de Astrobiolog\'{i}a, CAB, Spain),
Laura Colzi (Centro de Astrobiolog\'{i}a, CAB, Spain),
V\'ictor Rivilla (Centro de Astrobiolog\'{i}a, CAB, Spain),
Francisco Montenegro-Montes (Universidad Complutense de Madrid, UCM, Spain)
Miguel Sanz-Novo (Centro de Astrobiolog\'{i}a, CAB, Spain),
Marta Rey-Montejo (Centro de Astrobiolog\'{i}a, CAB, Spain),
Andr\'es Meg\'ias (Centro de Astrobiolog\'{i}a, CAB, Spain),
David San Andr\'es (Centro de Astrobiolog\'{i}a, CAB, Spain),
Sergio Mart\'in (European SOuthern Obseratory, ESO, Chile)
Shaoshan Zeng (RIKEN, Japan),
Amelie Godard (Centro de Astrobiolog\'{i}a, CAB, Spain),
Miguel Requena-Torres (Towson University, USA)
Juris Kalvāns (Ventspils University of Applied Sciences, Latvia)
\newline


\textbf{Science Keywords:} 
ISM: bubbles; ISM: molecules; ISM: clouds. 
\linebreak
\newline
\vspace{15mm}

 \captionsetup{labelformat=empty}
\begin{figure}[h]
   \centering
\includegraphics[width=\textwidth]{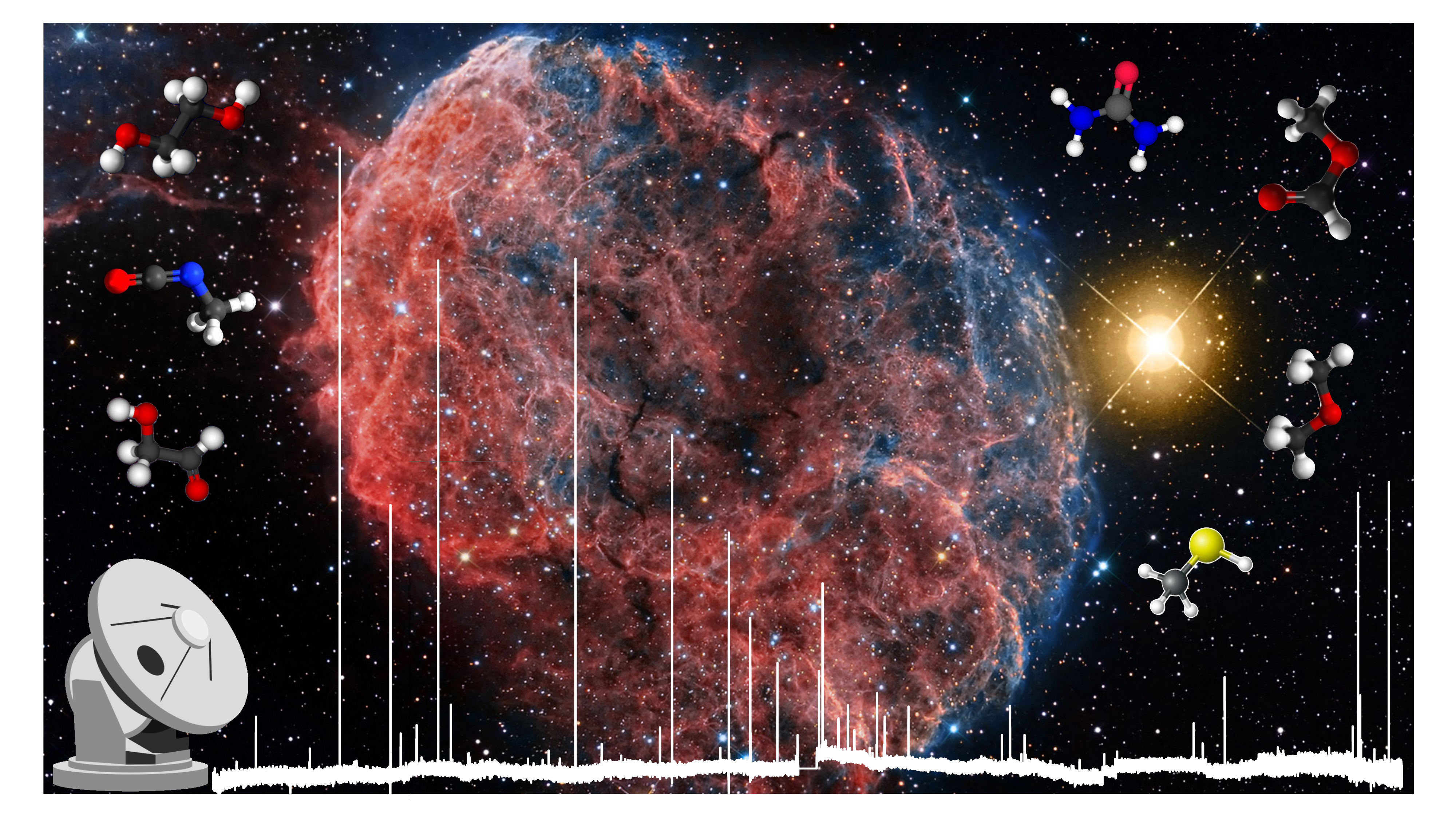}
   \caption{}
\end{figure}
\vspace{-15mm}

\setcounter{figure}{0}
\captionsetup{labelformat=default}


\pagebreak

\justifying
\section*{Abstract}
There is growing evidence that the Sun might have formed within a nebula impacted by at least one SNR. In this scenario, ejecta and shocks from SNRs may have provided the elements on which life as we know it is based. Investigating the chemical complexity of molecular clouds impacted by SNRs is therefore essential to unveil the star formation process and how life appeared on Earth. In this paper, we exploit this scientific questions and describe which technical specifications will drive in future generation telescopes. 

\section{Scientific context and motivation}
\textit{How did life come to be on Earth?} The answer to this fascinating and fundamental question remains one of the most elusive in both Astrophysics, Astrochemistry and Astrobiology. It is nowadays known that life on Earth emerged from the synthesis of simple compounds, known as pre-biotic and complex organic molecules (COMs), present in aqueous environments. From a chemical and biological point of view, pre-biotic molecules represent the building blocks of more complex compounds such as aminoacids, biopolymers (RNA and DNA), phospholipids (cellular membranes), which are at the very core of the human body functioning. Despite their crucial role of pre-biotic species and COMs, their origin and emergence on Earth is still unknown.
\newline

While some of these molecules might have formed in situ on the early Earth's surface, there is general consensus that a significant fraction of them was inherited by our planet from the proto-solar nebula and delivered to it during the Late Heavy Bombardment period. This hypothesis is supported by the recent discovery of four nucleobases and of 14 out of the 20 known proteinogenic amino acids in asteroids Ryugu and Bennu. However, which of these pre-biotic molecules and COMs were formed in the initial proto-solar nebula? Through which mechanisms? How complex these species can be? Recent studies on the chemical composition and complexity of primitive meteorites seem to indicate that supernova remnants (SNRs) might have played a significant role.

\vspace{-3mm}
\section{Science case}
SNRs are powerful feedback that profoundly affect the physical and chemical properties of the Interstellar medium (ISM; Chevalier et al. 1974, Slane et al. 2016, Trapp et al. 2024). At the typical cloud spatial scales (few parsecs), SNR-driven shocks can compress the surrounding material (Federrath et al. 2012), funnelling it into filaments (Appel et al. 2023) and setting the initial conditions of star formation (Inutsuka et al. 2015). In these high-density environments, fast gas cooling and enhanced chemical reactions can be enabled (Fragile et al. 2003).\\

\begin{figure}
    \centering
    \includegraphics[width=\linewidth,trim=11.8cm 2cm 6.9cm 0cm, clip=True]{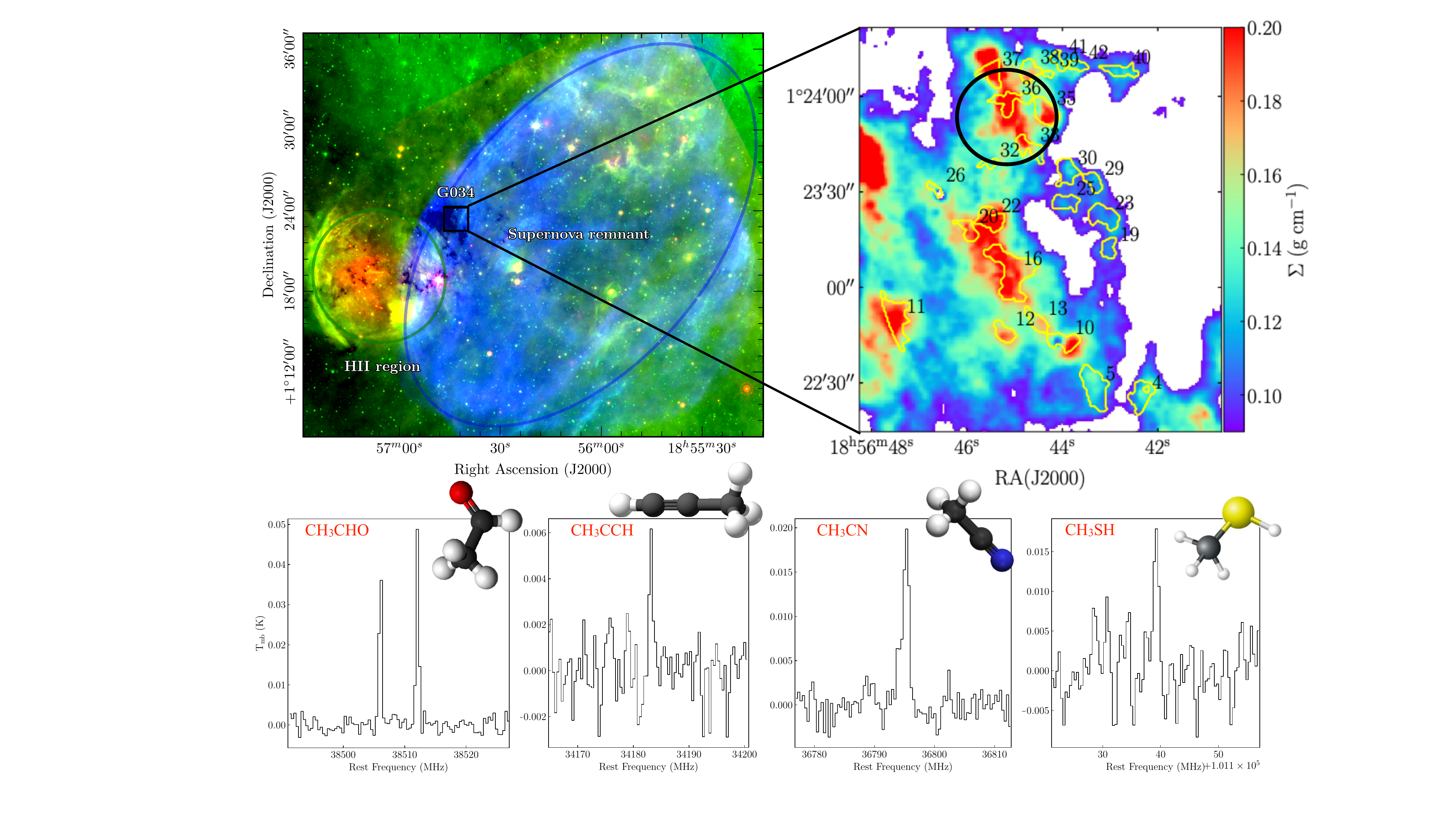}
    \caption{\textit{Top Left:} Three colour image of G34.77-00.55. Red is 24 $\mu$m emission from the MIPSGAL survey (Carey et al. 2009), green is 8 $\mu$m emission from the GLIMPSE survey (Churchwell et al. 2009) and blue is 1 GHz continuum emission from the THOR survey (Beuther et al. 2016). \textit{Top Right:} Mass surface density map (Kainulainen \& Tan 2013) with indicated the location of the triggered star formation (cores; Cosentino et al. 2023a). \textit{Bottom:} Rotational transitions of COMs identified toward a region of triggered star formation (Cosentino et al. 2025d sub.)}
    \label{fig:fig1}
\end{figure}

Recently, it has been suggested that the Sun itself might have formed through this mechanism, i.e., within the protosolar nebula impacted by at least one SNR (Bosset al. 2013; Young et al. 2014; Parker et al. 2023). According to this scenario, a SNR-driven shock might have caused the protosolar nebula to collapse, simultaneously enriching its material with radioactive and chemically processed elements from the ejecta (e.g. Boss \& Keiser 2010). Short-lived radionuclides, such as $^{26}$Al and $^{60}$Fe, found in primitive meteorites seems to supports this scenario (Desch et al. 2024). Recent high-angular resolution studies carried out with ALMA have investigated in detail the physical properties of the post-shocked gas at the interface between SNRs and molecular clouds (Fig.~\ref{fig:fig1}; Cosentino et al. 2019). The shock compresses the molecular gas to densities high enough for dense cores to start collapsing, as revealed by the enhancement of deuterated species such as N$_2$D$^+$ typical of low-mass prestellar cores (Cosentino et al. 2023). Follow-up high-sensitivity, single pointing observations demonstrate that simple, sub-thermally excited COMs such as CH$_3$OH, CH$_3$CN, CH$_3$CHO and CH$_3$SH are already present in these dense cores, indicating that chemical complexity starts early in the process of star and planet formation induced by SNRs (Cosentino et al. 2025d). 

Despite the latest advancement, current mm and sub-mm facilities do not have the technical capabilities necessary to comprehensively investigate the role of SNRs in enabling pre-biotic chemistry in molecular clouds. State-of-the-art single dish facilities such us IRAM-30m, Yebes-40m have been enabling high-sensitivity, broad bandwidth spectral survey at multiple wavelengths (Cosentino et al. 2023, 2025a, 2025d). However, to achieve the requested sensitivities (few mK) toward the large areas in the sky typical of SNRs results in unrealistic integration times. This is even more true for world-leading interferometric facilities such as ALMA and NOEMA, for which a very large amount of pointings would be required. As an example, the IRAM-30m and Yebes-340m telescopes require an integration time of 20-30 hours per pointing, to achieve a sensitivity of 3 mK at 3 and 7 mm, respectively. That is considering a moderate spectral resolution of 200 KHz. Similarly, to cover the large spatial extent typical of SNRs such as IC443 (30$^{\prime}\times$30$^{\prime}$), requires more than 3000 ALMA pointings (12m array) at 3 mm. So far, this limitation has been overcame by targeting specific regions of SNRs where shocks are already known to be present. Finally, the current achievable sensitivity, even in single spectra, are not sufficient to detect the larger COMs, especially those of prebiotic interest. Therefore, {\it ultrasensitive spectroscopic surveys of large portion of the sky are crucial to understand whether the special chemistry that led to the origin of life is a natural outcome of the star formation process induced by the interaction with SNRs.}  

\section{Technical requirements}
In light of all this, the science case here described can only be achieved by a facility that provides \textit{simultaneously} ultra-high spectral sensitivity (<1 mJy/beam) throughout a broad spectral bandwidth and for a very large fields of view. The Atacama Large Aperture Submillimeter Telescope (AtLAST) is the only upcoming facility able to provide all these capabilities at once. 
\newline

Expected in mid-2030, AtLAST will be a 50 m diameter single-dish facility located in proximity of the ALMA site (see Mroczkowski et al. 2025 for details). It will operate at frequencies between $\sim$ 30 and 950 GHz (0.3-10mm), covering a bandwidth that is almost comparable to that of the IRAM-30m, Yebes-40m and APEX telescope combined. Since pre-biotic molecules and COMs are expected to be sub-thermally excited, observations in the ALMA Band 1 frequency range (30-50 GHz) are essential to target bright low-J transitions. However, the simultaneous broad bandwidth of AtLAST will allow to cover not only the 7mm wavelength range, but also higher frequencies at 3mm, 2mm and 1mm, which are needed to accurately constrain the excitation conditions not only of prebiotic molecules. Thanks to the planned AtLAST's multi-beam array technology, a simultaneous field of view of up to 2$^{\circ}$ is expected. This is more than sufficient to cover the entire extent of most Galactic SNRs, or even several at once. Moreover, for the science case here described, high-sensitivities of $\sim$1 mJy/beam are necessary to detect very complex species. Since the observed line profiles can be as narrow as 1-2 km s$^{-1}$, high-spectral resolutions of $\sim$0.2 km s$^{-1}$ are also requested. 
\newline

With the requested specifications and assuming average weather conditions, AtLAST will require less than 50 hours to image extended SNRs at multiple wavelengths, while achieving a sensitivity of 1 mJy/beam at 35 GHz. For comparison, this is the amount of time that a single pointing with the Yebes 40m telescope would require to achieve similar sensitivity and spectral resolution.   
\newline

\noindent
\textbf{References.} Appel et al. 2023, ApJ, 954, 93 $\bullet$
Beuther et al. 2016, A\&A, 595, A32 $\bullet$
Boss \& Keiser 2010, ApJL, 717, L1 $\bullet$
Boss et al. 2013, ApJ, 76, 72 $\bullet$
Carey et al. 2009,  PASP, 121, 76 $\bullet$
Chevalier et al. 1974, , ApJ, 188, 501 $\bullet$
Churchwell et al. 2009, PASP, 121, 213 $\bullet$
Cosentino et al. 2019, ApJ, 881, L42 $\bullet$
Cosentino et al. 2023, A\&A, 675, A190 $\bullet$
Cosentino et al. 2025a, A\&A, 693, A199 $\bullet$
Cosentino et al. 2025d, arXiv:2512.07562 $\bullet$
Desch et al. 2024, Protostars and Planets VII, 759 $\bullet$
Federrath et al. 2012, ApJ, 761, 156 $\bullet$
Fragile et al. 2003, ApJ, 590, 778 $\bullet$
Inutsuka et al. 2015, A\&A, 580, A49 $\bullet$
Kainulainen \& Tan 20133, A\&A, 549, A53 $\bullet$
Mroczkowski et al. 2025,  A\&A, 694, A142 $\bullet$
Parker et al. 2023, MNRAS, 521, 4838 $\bullet$
Slane et al. 2016, Supernova Remnants Interacting with Molecular Clouds: X-Ray and Gamma-Ray Sig-
natures, ed. A. Balogh, A. Bykov, J. Eastwood, \& J. Kaastra, Vol. 51, 187  $\bullet$
Trapp et al. 2024, MNRAS, 533, 3008 $\bullet$
Young et al. 2014, Earth and Planetary Science Letters, 392, 16
\newline

\noindent
\textbf{Cover figure credits:} Velimir Popov \& Emil Ivanov 2020, Cosentino et al. 2025d, submitted.

\end{document}